\documentclass[a4paper,11pt]{article}
\usepackage{pos}
\usepackage{graphicx}
\usepackage{subfig}
\usepackage{lineno}
\usepackage{booktabs}
\usepackage{mathtools}
\usepackage{siunitx}
\title{The Acoustic Module for the IceCube Upgrade}
\author{The IceCube Collaboration \\{\normalsize \normalfont(a complete list of authors can be found at the end of the proceedings)}}

\emailAdd{cguenther@physik.rwth-aachen.de}
\emailAdd{juergen.borowka@rwth-aachen.de}

\abstract{The IceCube Neutrino Observatory will be upgraded with more than 700 additional optical sensor modules and new calibration devices. Improved calibration will enhance IceCube’s physics capabilities both at low and high neutrino energies. An important ingredient for good angular resolution of the observatory is precise calibration of the positions of optical sensors. Ten acoustic modules, which are capable of receiving and transmitting acoustic signals, will be attached to the strings. These signals can additionally be detected by compact acoustic sensors inside some of the optical sensor modules. With this system we aim for calibration of the detectors’ geometry with a precision better than 10 cm by means of trilateration of the propagation times of acoustic signals. This new method will allow for an improved and complementary geometry calibration with respect to previously used methods based on optical flashers and drill logging data. The longer attenuation length of sound compared to light makes the acoustic module a promising candidate for IceCube-Gen2, which may have optical sensors on strings with twice the current spacing. We present an overview of the technical design and tests of the system as well as analytical methods for determining the propagation times of the acoustic signals.

\vspace{4mm}
{\bfseries Corresponding authors:}
J\"urgen Borowka$^{1*}$, Christoph G\"unther$^{1*}$
Dirk Heinen$^{1}$, Simon Zierke$^{1}$\\
{$^{1}$ \itshape III. Physikalisches Institut B, RWTH Aachen University, D-52056 Aachen, Germany}\\[4mm]
$^*$ Presenter\\
\FullConference{37$^{\rm{th}}$ International Cosmic Ray Conference (ICRC 2021)\\
		July 12th -- 23rd, 2021\\
		Online -- Berlin, Germany}
}

\begin{document}
\maketitle

% \section{Introduction}

% The IceCube neutrino telescope \cite{Aartsen:2016nxy} is a cubic-kilometer size detector at the geographic South Pole that instruments the Antarctic ice in a depth between 1.5\,km and 2.5\,km. Charged particles from neutrino interactions produce Cherenkov light in the ice, which is detected by so-called Digital Optical Modules (DOMs) to reconstruct the neutrino's energy and direction. Crucial for a precise directional reconstruction is the knowledge of the properties of the optical medium as well as the location of the DOMs, ideally to an accuracy better than $1\,ns \times c_{\mathrm{ice}} \simeq 20\,\mathrm{cm}$. Currently an uncertainty of the DOM position of 50\,cm to 100\,cm has to be assumed depending on the depth and drilling procedure. This contributes about 10\,\% to the total directional uncertainty of reconstructed high-energy muon tracks \cite{Lilly2019}.

\section{Introduction}
With the IceCube Upgrade \cite{Ishihara:2019uL} approximately 700 new optical modules and multiple new calibration devices %(see e.g. \cite{Fruck:2019iK}, \cite{Kang:2019n+}, \cite{Ishihara:201999}) 
will be deployed at the center of the existing IceCube neutrino telescope \cite{Aartsen:2016nxy}. These will be mounted on 7 new strings with a horizontal spacing of 30\,m and 3\,m vertically between modules along the strings. The goal of the calibration devices is to improve the understanding of the ice properties and to test new methods for the geometrical calibration of the detector. The precise knowledge of the positions of the Digital Optical Modules (DOMs) to an accuracy better than $1\,\mathrm{ns} \times c_{\mathrm{ice}} \simeq 20\,\mathrm{cm}$ will significantly reduce the directional uncertainty of reconstructed high-energy muon tracks \cite{Lilly2019}.

The acoustic module (AM) presented here benefits from the results of the successful measurements of the South Pole Acoustic Test Setup (SPATS). This was an array of piezo transducers deployed in 2007 at the South Pole Station to measure the acoustic properties relevant for the acoustic detection of astrophysical neutrinos \cite{Abbasi_2010}.

The AM is a standalone calibration device capable of emitting and receiving acoustic signals. This allows the propagation times of acoustic signals between them to be measured. By trilateration the system will deliver an improved and complementary geometry calibration in addition to methods based on optical propagation time and drill logging data. Due to the longer attenuation length of sound compared to light in ice, the optical calibration method can be improved on larger scales by cross-calibration. This enables calibrating the full IceCube detector, allowing reprocessing of already taken data. Moreover, the longer range of signals makes acoustic calibration a promising method for the planned IceCube-Gen2 detector, which will have larger average string spacings of 240\,m \cite{Aartsen_2021}.
Since the acoustic measurements are optically dark, they can be performed in parallel to the normal detector operation and therefore propagation time measurements can be performed with a large repetition, improving signal to noise.
%However, the effect on other devices due to EMI has to be investigated.

By detecting transient acoustic signals, the acoustic system may enable glaciological measurements such as studying the long-time movement of the ice or the dependence of the speed of sound on depth and direction. Finally, the continuous recording of acoustic signals allows for searching for acoustic signals in coincidence with optically detected high-energy neutrinos. A detailed study of the feasibility of acoustic neutrino detection based on SPATS data is given in \cite{Abbasi_2012}.

% \newpage
\section{Design of the Acoustic Calibration System}

\subsection{Overall System}
\begin{figure}[]
    \centering
    % \subfloat[\centering Footprint of the new strings within the IceCube detector.]{{\includegraphics[width=6cm]{Figures/Upgrade_Strings_top_view.png}}}%
    % \qquad
    \subfloat[\centering]{{\includegraphics[width=7.1cm]{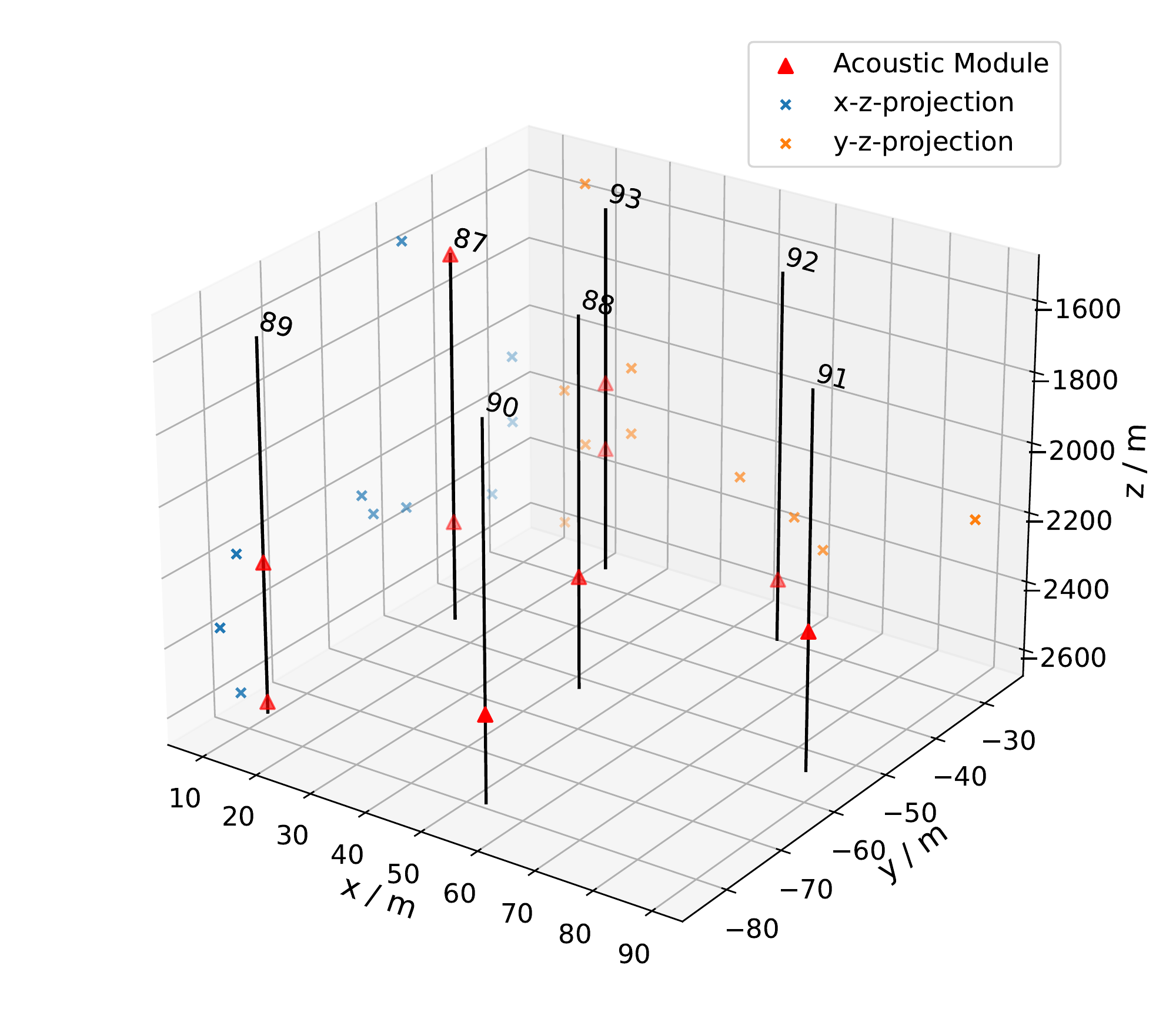}}\label{fig:upgradegeometryA}}
    \qquad
    \subfloat[\centering]{{\includegraphics[width=7.1cm]{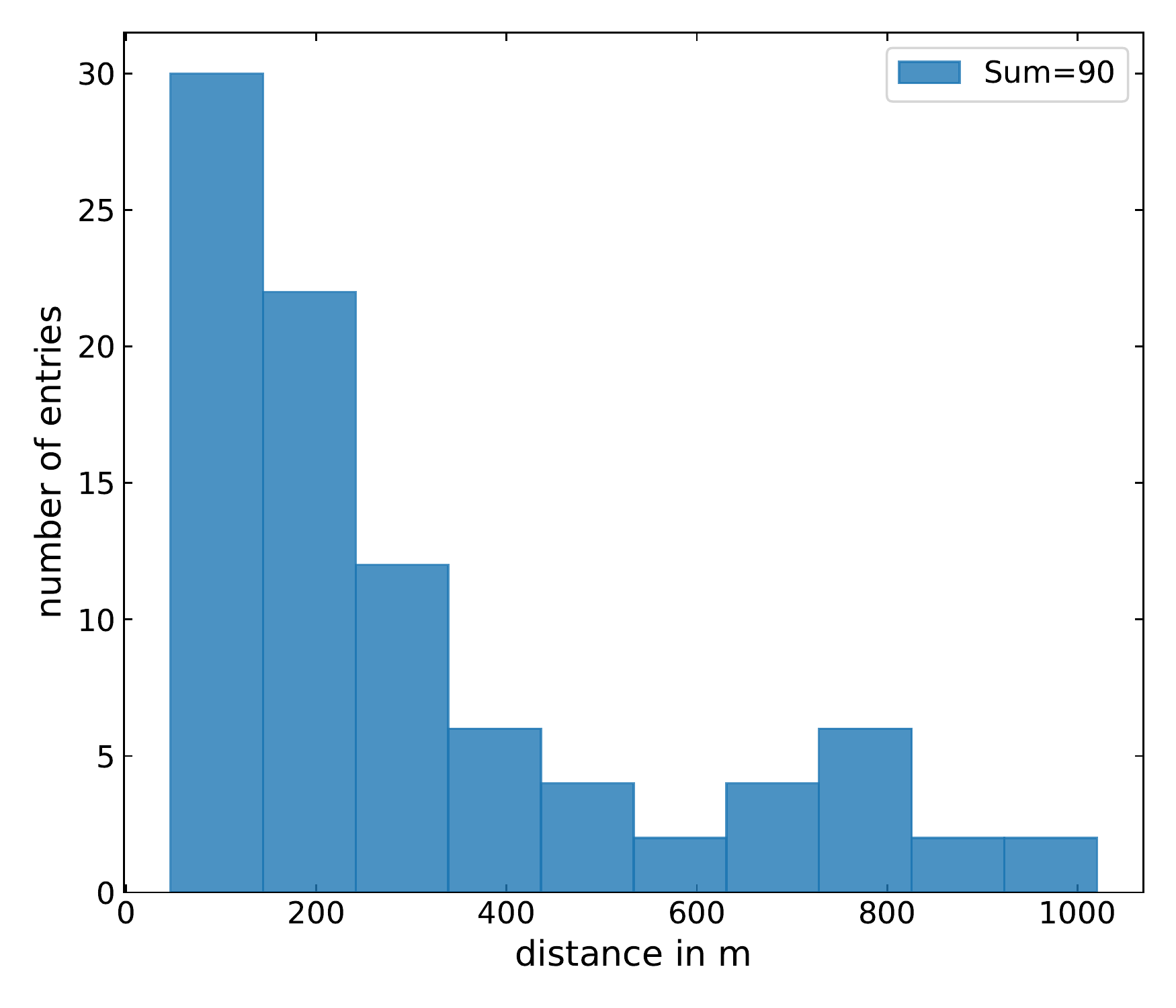}}\label{fig:upgradegeometryB}}
    \caption{\textbf{(a)} 3D view of the positions of the acoustic modules on the upgrade strings. \textbf{(b)} Histogram of the distances between acoustic modules. Distances are measured twice, in both directions.}
    \label{fig:upgradegeometry}%
\end{figure}

The technology of the acoustic calibration system is based on experience with the EnEx-RANGE project, in which an acoustic localization system for melting probes had been developed \cite{EnEx_2020}. The system
for the IceCube Upgrade consists of 10 AMs distributed over the new strings, as can be seen in figure \protect\ref{fig:upgradegeometryA}. Seven of the AMs will be located in the so-called physics region between a depth of 2140\,m to 2440\,m. To measure the sound propagation at large distances, three AMs will be placed outside this region, one of them at a significantly lower depth. This gives a wide distribution of distances between the AMs, up to $\simeq1000\,$m, as shown in figure \protect\ref{fig:upgradegeometryB}.

In addition to the AMs, acoustic receivers will be integrated into some optical modules, namely the pDOM \cite{DuVernois:201613}. An overview of these acoustic sensors can be found in \cite{Shefali:2019rR}.

% \newpage
\subsection{Design of the Acoustic Modules}
\begin{figure}[]
    \centering
    \subfloat[\centering]{{\includegraphics[height=6.7cm]{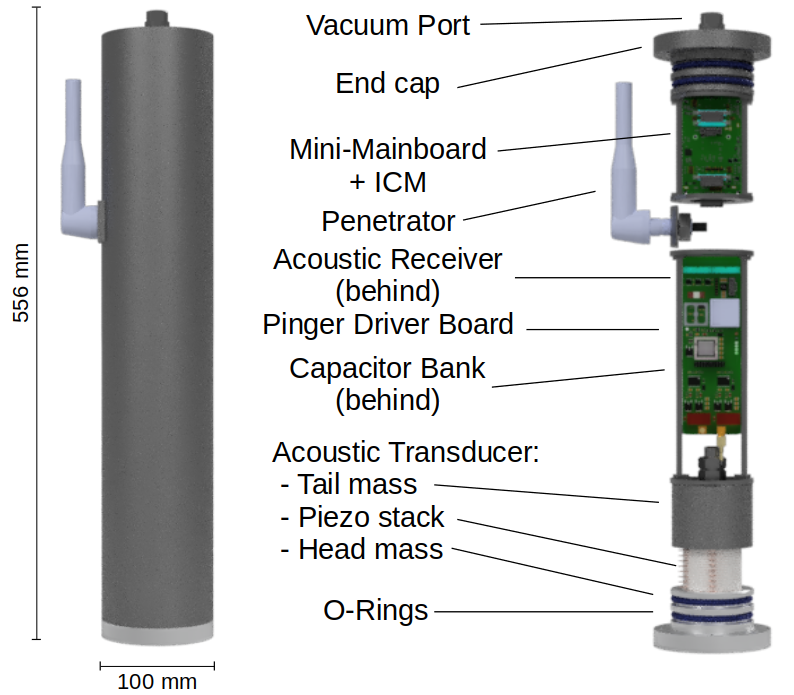}}\label{fig:AM_internal}}%
    \qquad
    \subfloat[\centering]{{\includegraphics[height=6.7cm]{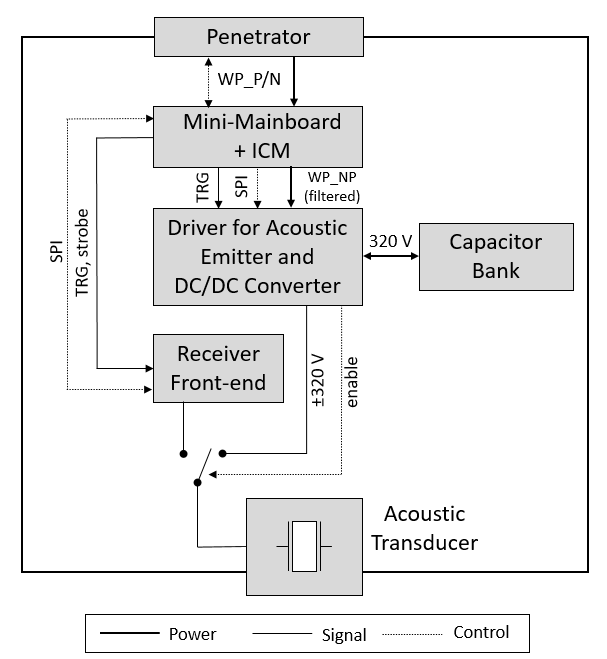}}\label{fig:AMblockdiagram}}%
    % \qquad
    % \subfloat[\centering Picture of the acoustic module prototype. (Placeholder)]{{\includegraphics[width=7cm]{Figures/AM_Foto.jpg}}}%
    \caption{\textbf{(a)} Illustration of the acoustic module showing its internal components. \textbf{(b)} Block diagram of the acoustic module components.}%
    \label{fig:AMcomponents}%
\end{figure}

An illustration of the AM and its components and their interconnection is shown in figure \ref{fig:AMcomponents}. The pressure housing is made mainly of steel and is designed to withstand pressures of up to 70\,MPa. The housing consists of a cylinder enclosed by two caps at the ends that are sealed with O-rings. The housing has a total length of 556\,mm, an outer (inner) diameter of 100\,mm (71\,mm) and a mass of $\approx 23$\,kg. A vacuum port allows the pressure inside the housing to be regulated. Cables are led into the housing by a penetrator cable assembly (PCA) that is mounted at the side. % to avoid bending the cable.

The acoustic module is powered by a 96\,V differential wire pair (WP), which is connected to the Mini-Mainboard (MMB). This board is used by many devices in the IceCube Upgrade and filters the communication signals from the WP and creates a 5\,V power rail for supplying the other components.
%The supply voltage (96\,V) differential wire-pair signals (WP) of the PCA are connected to the Mini-Mainboard (MMB). This is used by multiple devices in the IceCube Upgrade and consists of two separate PCBs. The so-called MMB power board filters the communication signals from the wire-pairs and creates a 5\,V power rail for supplying other devices.
Communication to the surface and time synchronization is provided by the Ice-Comms-Module (ICM) used by all Upgrade devices. Through the ICM, commands can be sent to the MMB which hosts an STM32 microcontroller and controls the other electronic components.
%The interconnection of the components is shown in the block diagram in figure \ref{fig:AMblockdiagram}.

The Pinger Driver Board generates the high voltage signals to drive the acoustic transducer. A flyback DC/DC converter circuit charges a $\sim400$\,$\mu$F capacitor bank to 320\,V. The circuit is designed to consume $\simeq3\,$W during charging, resulting in charge times of approximately 30\,s. The output signals are generated with a full-bridge driver circuit that can switch the output voltage at 3 different voltage levels -320\,V, 0\,V, 320\,V with a sampling rate of up to $1\,$MSps. By varying the output state and duration, arbitrary waveforms such as sine sweeps (chirps) can be generated.

The acoustic transducer consists of a stack of 16 piezo ceramic discs (Sonox P4 from CeramTec) pre-stressed in between the tail mass and the head mass by a threaded rod (Tonpilz design). The discs have an outer (inner) diameter of 50\,mm (15\,mm) and a thickness of 2\,mm. The tail and head mass are made of steel and aluminium, respectively. This results in a mass ratio of $M_{\mathrm{head}} / M_{\mathrm{tail}} \approx 1:2$ increasing the acoustic power radiated outwards and setting a resonance frequency of the transducer of $\sim11\,$kHz. Aluminium is a favourable material for the head mass due to its
acoustic impedance %$Z=\rho\cdot c \simeq 17\times10^6\,$kg\,m$^{-2}$\,s$^{-1}$
, which increases the transmission coefficient for the transition from piezo ceramic to ice.

The AMs contain an acoustic receiver board that can digitize electric signals from the transducer at a sampling rate of up to 100\,kHz (see \cite{Shefali:2019rR} for more details). This circuit also is used for the acoustic sensors in the pDOMs.%To prevent damaging the receiver, relays allow either only the high-voltage signal generating circuit or the receiver to be connected to the transducer but not both.

% Components:
% \begin{itemize}
%     \item Pressure Hull (dimension, penetrator, vacuum port)
%     \item Mini-Mainboard (Controller + Power Board, ICM)
%     \item Pinger Frontend (HV supply (320V), signal generation, capacitor bank)
%     \item Acoustic receiver
%     \item Acoustic transducer (Tonpilz configuration, Piezos Sonox P4, head/tail mass, mass ratio, resonance frequency 10kHz (measurement))
% \end{itemize}

% \newpage
\section{Analysis Method and expected Performance}

\subsection{Propagation Time Measurement}
Two methods are common to determine the propagation time of acoustic signals in ice or water, the $5\sigma$-threshold \cite{Lars2019} and  the cross-correlation method \cite{Blip2007}. The former analyzes the receiver output signal in the time domain to detect first time when the signal is above a certain threshold, i.e. a level of 5 times the standard deviation $\sigma$ of the background noise. The latter calculates the cross-correlation product of emitter input signal with receiver output signal to determine the point in time when this product is at its maximum. A drawback of the threshold method is its sensitivity against any signal event rising above the threshold while the cross-correlation method needs a special signal type (e.g. chirp) as emitter input to guarantee unambiguous results. Both procedures have to be synchronized with the starting point in time, when the acoustic signal is generated.

\subsubsection{Phase Response Method}
In this paper we introduce a new robust and precise measurement procedure to determine the propagation time of acoustic signals in a media, called the phase response method. The method is based on aspects of system theory of linear time-invariant systems (LTI systems).

Two aspects are important for the application of the phase response method. The first one is the property of a subclass of LTI systems, namely the minimum-phase systems. These systems are the ones with the smallest possible group delay $\tau$, where group delay is defined as the negative derivative of the phase response with respect to angular frequency $\omega$. The second important aspect is the Bode gain-phase relation, sometimes also referred to as Bayard-Bode theorem.

The propagation time measurement chain consists of a sine sweep input signal $x(t)$ which is fed into an acoustic emitter. The generated acoustic signal travels through a medium (i.e. ice) and after a propagation time $\tau$ is measured by an acoustic receiver, giving the output signal $y(t)$. % Emitter and receiver are submerged into the acoustic media.

Such an LTI system can be described as a chain of three subsystems (emitter - acoustic medium – receiver), characterized either in the time domain by the convolution of the impulse responses or in the frequency domain by the product of the frequency responses of the subsystems. The frequency response of a LTI system is the Fourier transform of the system's impulse response.

\begin{figure}[]
    \centering
    \subfloat[\centering]{{\includegraphics[width=6.5cm]{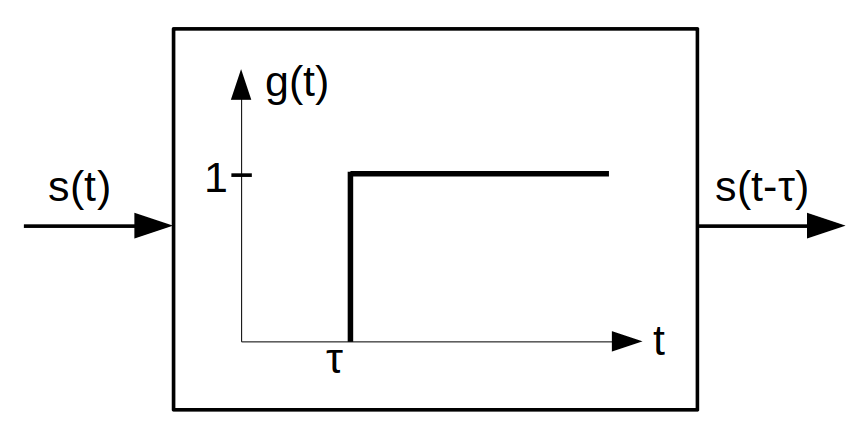}}\label{fig:juergen3a}}%
    \qquad
    \subfloat[\centering]{{\includegraphics[width=6.5cm]{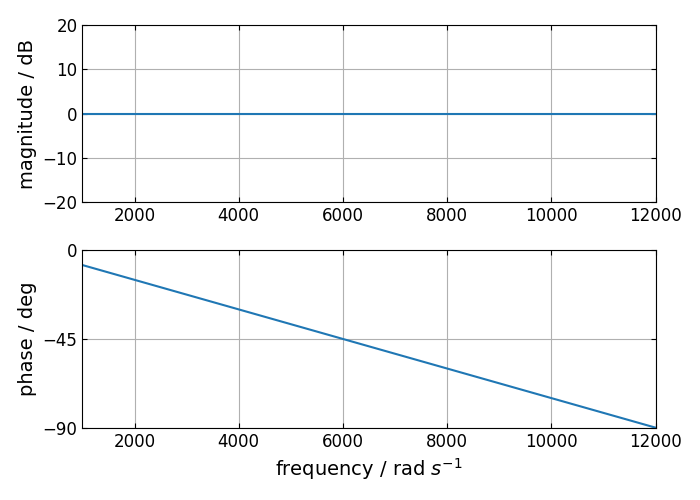}}\label{fig:juergen3b}}%
    \caption{\textbf{(a)} LTI system model (step response $g(t)$) of the acoustic medium as a delay time element. \textbf{(b)} Frequency response $H_{\tau}(\omega)$ of delay time element.}
    % \label{fig:AMcomponents}%
\end{figure}

The subsystem acoustic medium has only the property of a certain time delay $\tau$, which is the time the acoustic signal needs to travel in the medium between emitter and receiver (see figure \ref{fig:juergen3a}). Any attenuation of the signal can be described by multiplication  factor in the receiver subsystem. In the frequency domain such a delay time element is characterized by its frequency response $H_{\tau}(\omega) = \exp(-j\omega \tau)$. It has a constant magnitude of 1 over all frequencies and a linearly decreasing phase with a slope of $-\tau$ (see figure \ref{fig:juergen3b}).

Assuming the emitter and receiver can be considered to be minimum-phase systems\footnote{If this is not true, the phase response method still holds, but a few more considerations have to be taken into account.}, the in-between element delay time makes the whole system a non-minimum-phase system since a value of $\tau$ is added to the group delay. The magnitude $|H(\omega)|$ of the frequency response of the signal chain however remains the same when adding the delay time element. $H(\omega)$ can be written as:
\begin{equation}
    H(\omega) = |H(\omega)|\cdot e^{-j\omega\tau} = H_{\mathrm{emit}}(\omega) \cdot H_{\tau}(\omega) \cdot H_{\mathrm{rec}}(\omega) = \mathcal{F}\{y(t)\} / \mathcal{F}\{x(t)\}
\end{equation}
where $H_{\mathrm{emit}}(\omega)$ and $H_{\mathrm{rec}}(\omega)$ denote the frequency response of the emitter and receiver, respectively. $H(\omega)$ can thus be calculated from the ratio of the Fourier transforms of the measured signals $y(t)$ and $x(t)$. $H(\omega)$ is a complex valued function, with a magnitude $|H(\omega)|$ and a phase response $\phi(\omega)$.

The Bayard-Bode theorem states that for a minimum-phase system one can calculate the phase response given the magnitude of the frequency response. Since the delay time element does not affect $|H(\omega)|$:
\begin{equation}
    |H(\omega)| = |H_{\mathrm{emit}} \cdot H_{\mathrm{rec}}| \eqcolon |H_0(\omega)|
\end{equation}
The phase response $\phi_0(\omega)$ of $H_0(\omega)$ is calculated out of $|H_0(\omega)|$, using the Bayard-Bode algorithm  \cite{bayardbode1940}. Thus, finally:
\begin{equation}
    H(\omega) = H_0(\omega) \cdot H_{\tau}(\omega), \:\mathrm{with}\: \phi(\omega) = \phi_0(\omega) + \phi_{\tau}(\omega) \implies \tau = -\frac{d}{d\omega} (\phi(\omega) - \phi_0(\omega))
\end{equation}

\subsection{Simulation of the Array Performance}
The localization performance of the array of AMs has been estimated by the following procedure and assumptions. A more detailed description can be found in \cite{Max2020}.

\begin{figure}[]
    \centering
    \subfloat[\centering]{{\includegraphics[width=7cm]{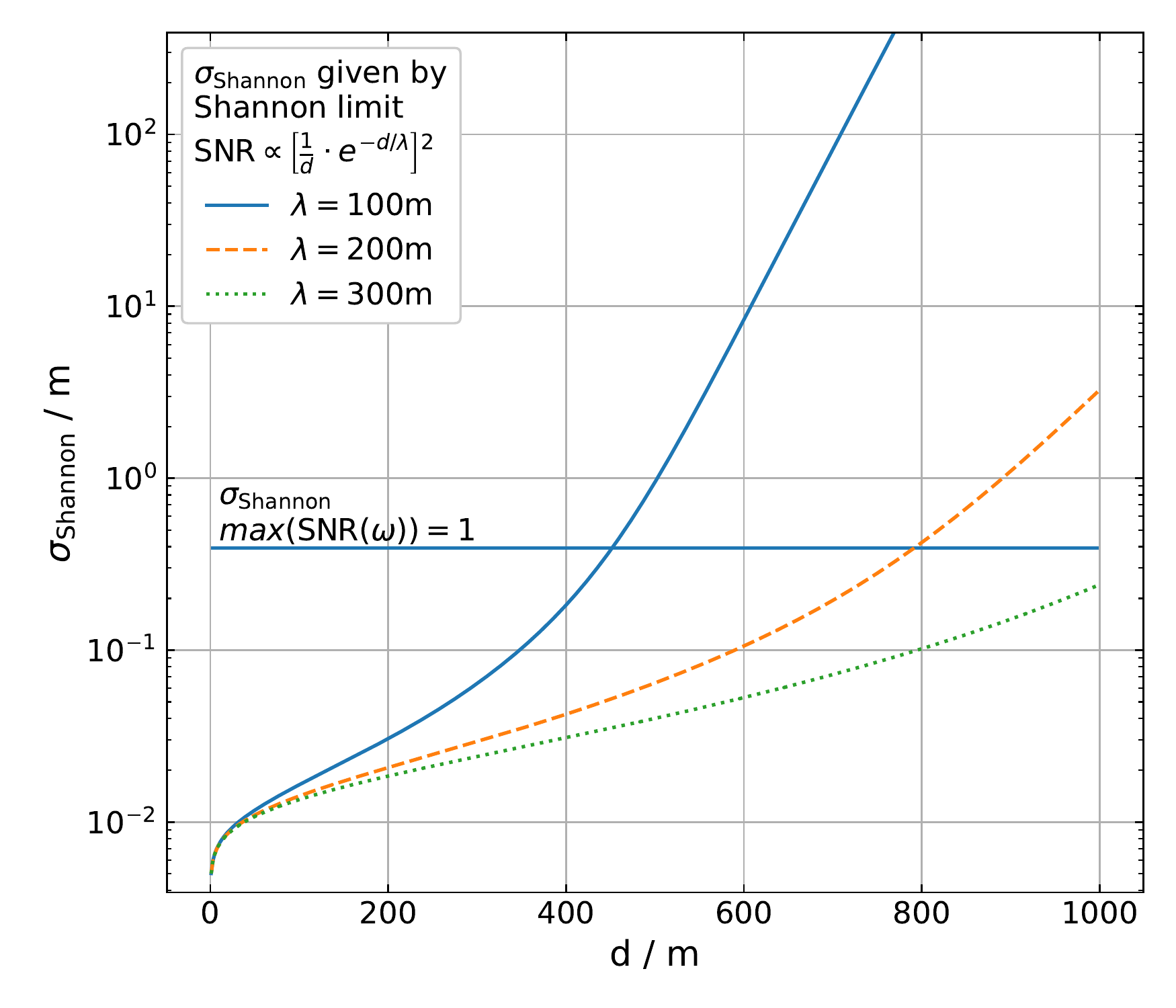}}\label{fig:shannon}}%
    \qquad
    \subfloat[\centering]{{\includegraphics[width=7cm]{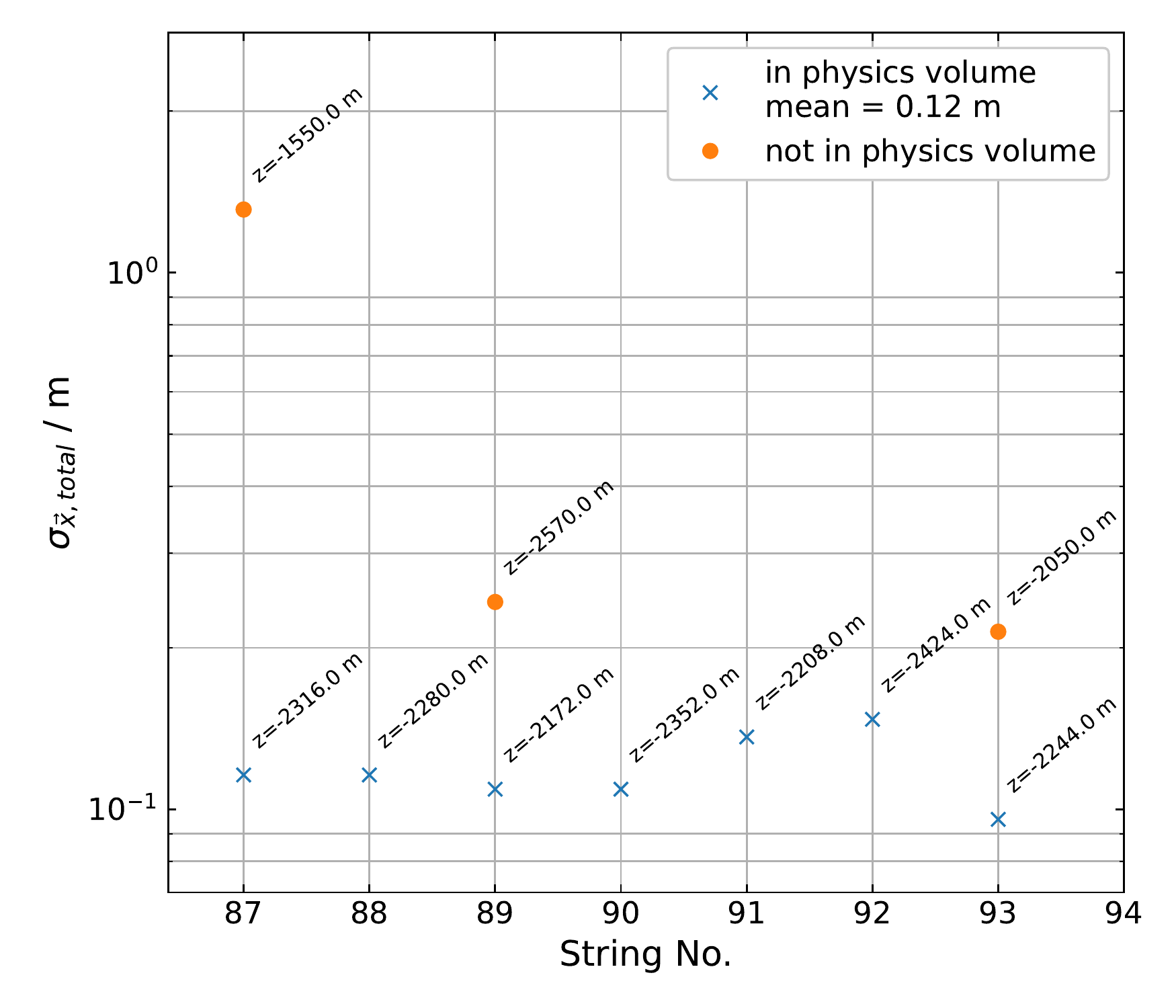}}\label{fig:alluncert}}%
    \qquad
    \subfloat[\centering]{{\includegraphics[width=7cm]{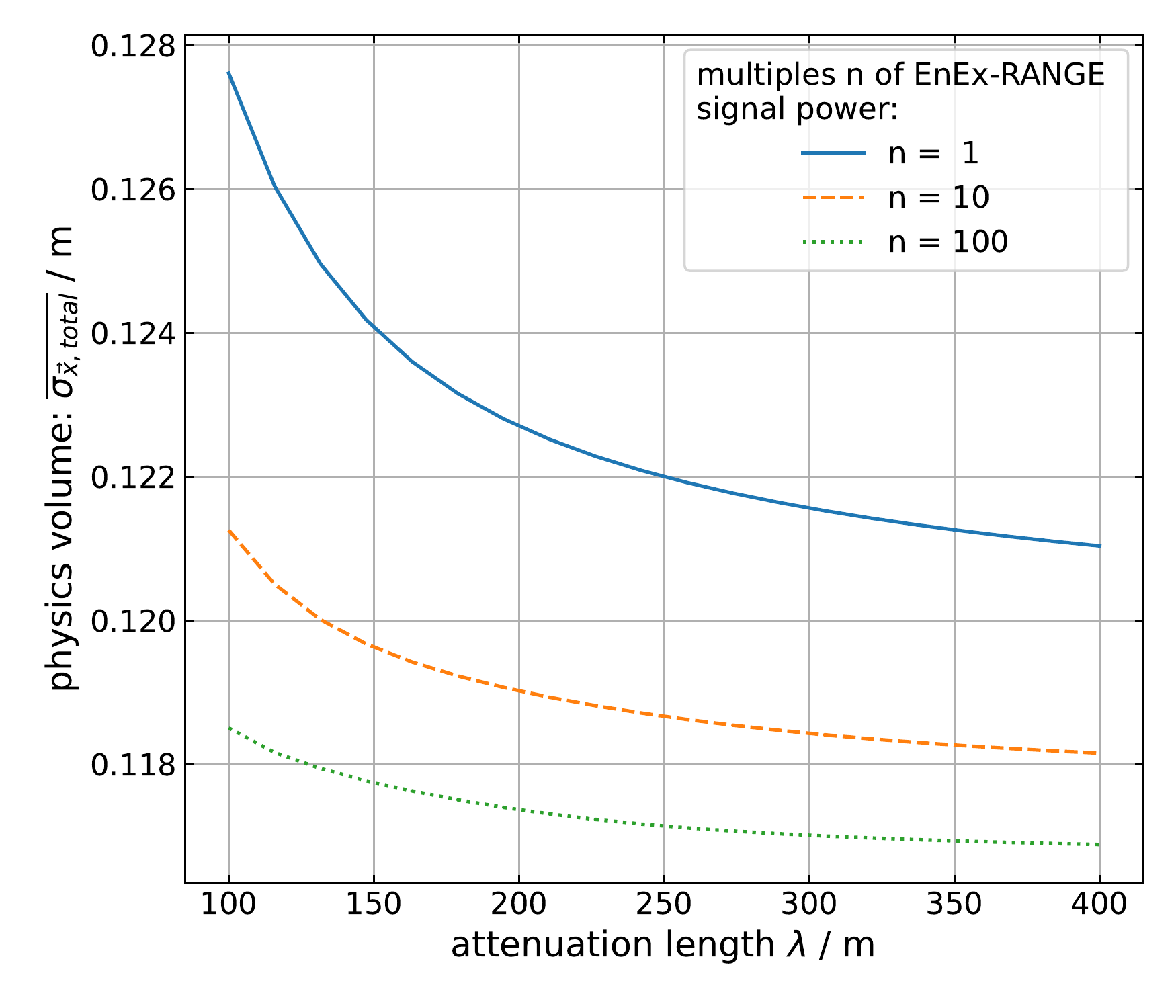}}\label{fig:uncertphys}}%
    \caption{\textbf{(a)} Distance dependence of $\sigma_{\mathrm{Shannon}}$. The horizontal line marks the distance at which SNR=1. \textbf{(b)} Calculated uncertainties of the individual acoustic modules. \textbf{(c)} Dependence of the average uncertainty on the attenuation length for different output powers for AMs in the physics volume.}%
    \label{fig:performanceplots}%
\end{figure}

For each given AM, N=9 different measurements of the propagation time to the respective other AMs can be performed. This amounts to a total of 90 times, where each path is measured twice in opposite directions. Different error sources contribute to the uncertainty of the propagation time measurement. The digitization uncertainty $\sigma_{\mathrm{digi}}\simeq10\,\mu$s results from the sampling frequency of the receiver and the synchronization uncertainty between the AMs. The size of the transducer head introduces a spatial uncertainty $\sigma_{\mathrm{spat}}\simeq 5\,$cm, which is estimated by half the outer diameter of the head mass.% The speed of sound is $c_{\mathrm{ice}} \simeq3900\,$m/s \cite{Abbasi_2010}. 
The uncertainty related to the distance-dependent accuracy of transmitted signal is determined by the signal relative to
the background noise and can be 
 estimated by the Shannon-Hartley theorem \cite{1697831}. There,
the maximum rate of information $C$ in bits per second depends on the signal-to-noise-ratio (SNR):
%\begin{equation}
$
    C = B \cdot \sum_{\omega} \log_2 \left( 1 + \mathrm{SNR}(r,\omega) \right)
    \label{eq:shannon}
$
%\end{equation}
, where the summation goes over all frequencies within the bandwidth $B$ of the signal.
% The minimum time interval $\sigma_{\mathrm{Shannon}}$ that can be resolved with a propagation time measurement is then given by:
% \begin{equation}
%     \sigma_{\mathrm{Shannon}} = \frac{1\,\mathrm{bit}}{C}
%     \label{eq:shannonuncert}
% \end{equation}
The signal amplitude depends on the geometric distance with a $1/r$ dependence and an exponential attenuation with the attenuation length $\lambda$. Therefore,
$ %\begin{equation}
    \mathrm{SNR}(r,\omega) = \frac{Y^2(\omega)}{N^2(\omega)} \cdot \left( \frac{1}{r} \cdot e^{-r/\lambda} \right)^2
    \label{eq:snrdep}
$
%\end{equation}
where $Y^2(\omega)$ is the frequency-dependent system answer\footnote{The system answer is described by the convolution of the excitation signal and the impulse response of all components of the signal path} and $N^2(\omega)$ is the power of the background noise. The acoustic attenuation in Antarctic ice at depths down to $500$\,m has been measured by the SPATS experiment \cite{Abbasi_2011} %the South Pole Acoustic Test Setup (SPATS) 
to be independent of frequency $\lambda = 300$\,m$\,\pm\,20\,\%$. The measurement of the speed of sound yields $c_{\mathrm{ice}} \simeq3900\,$m/s \cite{Abbasi_2010}.
Putting things together, %Using equation \ref{eq:shannon} and \ref{eq:snrdep} 
the minimum resolvable propagation time  $\sigma_{\mathrm{Shannon}}$ is given by:
\begin{equation}
%original  \sigma_{\mathrm{Shannon}} = \frac{1\,\mathrm{bit}}{C} = \left( \sum_{\omega} B \cdot \log_2 \left( 1 + \frac{Y^2(\omega)}{c} \left( \frac{1}{r} \cdot e^{-r/\lambda} \right)^2 \right) \right)^{-1}
    \sigma_{\mathrm{Shannon}} = \frac{1\,\mathrm{bit}}{C} = \left( \sum_{\omega} B \cdot \log_2 \left( 1 + n\cdot K_0 \left( \frac{1}{r} \cdot e^{-r/\lambda} \right)^2 \right) \right)^{-1}
    \label{eq:fullshannon}
\end{equation}
% Using equation \ref{eq:snrdep} into equation \ref{eq:shannonuncert} yields:
% \begin{equation}
%     \sigma_{\mathrm{Shannon}} = \left( \sum_{\omega} B \cdot \log_2 \left( 1 + \frac{Y^2(\omega)}{c} \left( \frac{1}{r} \cdot e^{-r/\lambda} \right)^2 \right) \right)^{-1}
%     \label{eq:fullshannon}
% \end{equation}
The absolute value of the SNR for the AM is still unknown (as is the attenuation length $\lambda$). Therefore, we replace the SNR by a scaling factor $n\cdot K_0 $. Here $K_0 = \SI{95e3}{m} $  corresponds to the  SNR  from the previous known EnEx-RANGE emitter-receiver-system that resulted in   $SNR = 25:1$ measured at a distance of 38\,m on a temperate alpine glacier of $\lambda=8.25\,$m attenuation
 \cite{EnEx_2020}. After distance and attenuation correction, this defines $K_0$ and the baseline is $n=1$. Due to the larger piezo stack (twice the number of discs) and improved head-to-tail mass ratio of the AM, a factor of ten higher emitted amplitude, $n=10$,  could be assumed, not yet including the potentially much higher number of repetitions, by which one may reach $n=100$ or more.
Figure \ref{fig:shannon} shows the distance dependence of $\sigma_{\mathrm{Shannon}}$ for different attenuation lengths. For small distances the uncertainty becomes zero and for large distances it follows an exponential law.

The total time uncertainty for each receiver-emitter pair is obtained by adding the above 
uncertainties in quadrature $\sigma_t^2 =
\sigma_{\mathrm{spat}}^2 + \sigma_{\mathrm{digi}}^2 + \sigma_{\mathrm{Shannon}}^2$.
Given the speed of sound $c_{ice}$ and $N$ time measurements, we can calculate for each receiver  
that displacement $\Delta \vec{x}$  from its nominal position that leads to a 
$\Delta \chi^2 =1 $, where $\chi^2 = \sum_N 
\left ( \frac{\Delta x }{c_{ice}\cdot \sigma_t} \right )^2 $. Repeating this procedure for every spatial coordinate, we can estimate the total localization uncertainty by $\sigma_{\vec{x},\mathrm{tot}}^2 =
(\Delta x )^2 + (\Delta y)^2 + (\Delta z)^2  $
for each receiver.
%
%  Given $N$ time measurements, the uncertainties $\sigma_{x_n}$ of the individual measurements are combined by
% \begin{equation}
% \frac{1}{\sigma_x^2} = \sum_{n}^N \frac{1}{\sigma_{x_n}^2}
% \label{eq:error}
% \end{equation}
% To simulate the effect of the propagation time uncertainty on the spatial uncertainty, the AMs are displaced by a small amount, so that the relation
% \begin{equation}
%     \frac{\sigma_{t_n}}{\sigma_{x_n}} \approx \frac{\overline{\tau}(x_n) - \overline{\tau}(x_n+\Delta x) }{|\Delta x|}
% \end{equation}
% holds. This way we can estimate the right side of equation \ref{eq:error} by
% \begin{equation}
%     \frac{1}{\sigma_x^2} = \sum_n^N \frac{1}{\sigma_{x_n}^2} \approx \sum_n^N \frac{\left( \overline{\tau}(x_n) - \overline{\tau}(x_n+\Delta x) \right)^2 }{\sigma_{t_n} (\Delta x)^2} = \frac{\chi^2(\Delta x)}{(\Delta x)^2}
% \end{equation}
% This means the total variance of a given AM can be calculated by computing the $\chi^2$ of a small displacement of the AM around its start position. This procedure can be applied to all three dimensions of the system resulting in
% \begin{equation}
%     \sigma_{x,\mathrm{tot}}^2 \approx \sum_{\Vec{e}\in[\Vec{x},\Vec{y},\Vec{z}]} \frac{(\Delta \Vec{e})^2}{\chi^2(\Delta \Vec{e})}
% \end{equation}
Figure \ref{fig:alluncert} shows the result of the estimated uncertainty for each AMs. The uncertainty of the AMs within the physics volume is on average $\sim12$\,cm. For the three AMs outside the physics volume, the uncertainty is larger. Especially the solitary AM has a significantly larger uncertainty on the order of $\sim1$\,m.  The larger distance results in a lower SNR as well a smaller angular lever arm of probed distances.

%Figure \ref{fig:uncertphys} shows the dependence of the average uncertainty of the AMs in the physics volume for different simulated signal powers and attenuation length. It can be concluded that for various different assumptions the uncertainty in the order of $\sim10$\,cm. 

The attenuation length of the Antarctic ice may decrease with larger depth, as the temperature of the ice increases with depth % \cite{VANDENBROUCKE2009S164},
\cite{Price7844}. Therefore, the simulation has been repeated assuming smaller values of the attenuation length and also for different output powers of the AMs. The result for  AMs in the physics volume is shown in figure \ref{fig:uncertphys}. Even for largely different assumptions the mean uncertainty remains robust at about $\sim12$\,cm.
%The same is shown in figure \ref{fig:uncertphysSol} for the solitary AM. One can see that in this case the uncertainty is highly dominated by the attenuation of the sound signals.

\section{Conclusion and Outlook}
We have presented the technical design of a high-power acoustic transducer that will be part of an acoustic calibration system in the IceCube Upgrade. Based on the theory of LTI systems, we introduced a robust and precise method for measuring the propagation time of acoustic signals in ice. Simulation results of the array of AMs show that excellent localization performance on the order of $\sim12\,$cm can be achieved. The exact performance still depends on some unknown ice parameters like the attenuation length of sound at depth below 2\,km, which is one of the expected results from the IceCube Upgrade. With these results the design can be further optimized for the upcoming IceCube-Gen2 detector. Due to the larger string spacings of $\simeq240\,$m, the AMs are a promising candidate for the geometric calibration of this future detector.

 \bibliographystyle{ICRC}
\bibliography{references}

% \newpage
% \begin{thebibliography}{99}
% \bibitem{...}
% ....

% \end{thebibliography}

% Full authors list (ONLY FOR COLLABORATIONS)
\clearpage
\section*{Full Author List: IceCube Collaboration}
\scriptsize
\noindent
R. Abbasi$^{17}$,
M. Ackermann$^{59}$,
J. Adams$^{18}$,
J. A. Aguilar$^{12}$,
M. Ahlers$^{22}$,
M. Ahrens$^{50}$,
C. Alispach$^{28}$,
A. A. Alves Jr.$^{31}$,
N. M. Amin$^{42}$,
R. An$^{14}$
% ...
\\

\noindent
% ...
$^{12}$ Universit{\'e} Libre de Bruxelles, Science Faculty CP230, B-1050 Brussels, Belgium \\
% ...
$^{14}$ Department of Physics and Laboratory for Particle Physics and Cosmology, Harvard University, Cambridge, MA 02138, USA \\
% ...
$^{17}$ Department of Physics, Loyola University Chicago, Chicago, IL 60660, USA \\
$^{18}$ Dept. of Physics and Astronomy, University of Canterbury, Private Bag 4800, Christchurch, New Zealand \\
% ...
$^{22}$ Niels Bohr Institute, University of Copenhagen, DK-2100 Copenhagen, Denmark \\
% ...
$^{28}$ D{\'e}partement de physique nucl{\'e}aire et corpusculaire, Universit{\'e} de Gen{\`e}ve, CH-1211 Gen{\`e}ve, Switzerland \\
% ...
$^{31}$ Karlsruhe Institute of Technology, Institute for Astroparticle Physics, D-76021 Karlsruhe, Germany  \\
% ...
$^{42}$ Bartol Research Institute and Dept. of Physics and Astronomy, University of Delaware, Newark, DE 19716, USA \\
% ...
$^{50}$ Oskar Klein Centre and Dept. of Physics, Stockholm University, SE-10691 Stockholm, Sweden \\
% ...
$^{59}$ DESY, D-15738 Zeuthen, Germany \\
% ...

\subsection*{Acknowledgements}

\noindent
USA {\textendash} U.S. National Science Foundation-Office of Polar Programs,
U.S. National Science Foundation-Physics Division,
U.S. National Science Foundation-EPSCoR,
Wisconsin Alumni Research Foundation,
Center for High Throughput Computing (CHTC) at the University of Wisconsin{\textendash}Madison,
Open Science Grid (OSG),
Extreme Science and Engineering Discovery Environment (XSEDE),
Frontera computing project at the Texas Advanced Computing Center,
U.S. Department of Energy-National Energy Research Scientific Computing Center,
Particle astrophysics research computing center at the University of Maryland,
Institute for Cyber-Enabled Research at Michigan State University,
and Astroparticle physics computational facility at Marquette University;
Belgium {\textendash} Funds for Scientific Research (FRS-FNRS and FWO),
FWO Odysseus and Big Science programmes,
and Belgian Federal Science Policy Office (Belspo);
Germany {\textendash} Bundesministerium f{\"u}r Bildung und Forschung (BMBF),
Deutsche Forschungsgemeinschaft (DFG),
Helmholtz Alliance for Astroparticle Physics (HAP),
Initiative and Networking Fund of the Helmholtz Association,
Deutsches Elektronen Synchrotron (DESY),
and High Performance Computing cluster of the RWTH Aachen;
Sweden {\textendash} Swedish Research Council,
Swedish Polar Research Secretariat,
Swedish National Infrastructure for Computing (SNIC),
and Knut and Alice Wallenberg Foundation;
Australia {\textendash} Australian Research Council;
Canada {\textendash} Natural Sciences and Engineering Research Council of Canada,
Calcul Qu{\'e}bec, Compute Ontario, Canada Foundation for Innovation, WestGrid, and Compute Canada;
Denmark {\textendash} Villum Fonden and Carlsberg Foundation;
New Zealand {\textendash} Marsden Fund;
Japan {\textendash} Japan Society for Promotion of Science (JSPS)
and Institute for Global Prominent Research (IGPR) of Chiba University;
Korea {\textendash} National Research Foundation of Korea (NRF);
Switzerland {\textendash} Swiss National Science Foundation (SNSF);
United Kingdom {\textendash} Department of Physics, University of Oxford.
\enddocument
\end{document}